\documentstyle[preprint,aps,epsfig]{revtex}
\tighten
\def\bea{\begin{eqnarray}}
\def\beann{\begin{eqnarray*}}
\def\beq{\begin{equation}}
\def\eea{\end{eqnarray}}
\def\eeann{\end{eqnarray*}}
\def\eeq{\end{equation}}
\def\nn{\nonumber}

\newcommand{\bcdot}{\bbox{\cdot}}

\newcommand{\bk}{\bbox{k}}
\newcommand{\bp}{\bbox{p}}
\newcommand{\bq}{\bbox{q}}

\newcommand{\bt}{\bbox{t}}

\newcommand{\bsigma}{\bbox{\sigma}}

\newcommand{\ra}{\!>}

\begin{document}
\headheight1.2cm
\headsep1.2cm
\baselineskip=20pt plus 1pt minus 1pt

\title{Quark-meson coupling model with constituent quarks:\\
Exchange and pionic effects}
\author{M.E. Bracco$^a$, G. Krein$^b$\thanks{\noindent
Alexander von Humboldt Research Fellow\hfil\break
Permanent Address: Instituto de F\'{\i}sica Te\'{o}rica, Universidade Estadual
Paulista\hfil\break
Rua Pamplona, 145 - 01405-900 S\~{a}o Paulo, SP - Brazil}, and M. Nielsen$^c$}
\address{$^a$ Instituto de F\'{\i}sica, Universidade do Estado do
Rio de Janeiro\\
Rua S\~{a}o Francisco Xavier 524, 20559-900, Rio de Janeiro, RJ - Brazil \\
$^b$ Institut f\"{u}r Kernphysik, Universit\"{a}t Mainz, 55099 
Mainz, Germany\\
$^c$ Instituto de F\'{\i}sica, Universidade de 
S\~{a}o Paulo, Caixa Postal 66318 \\
05315-970 S\~{a}o Paulo, SP - Brazil }
\maketitle
\begin{abstract}
The binding energy of nuclear matter including exchange and pionic
effects is calculated in a quark-meson coupling model with massive constituent
quarks. As~in~the case with elementary nucleons in QHD, exchange effects are
repulsive. However, the coupling of the mesons directly to the quarks in the 
nucleons introduces a new effect on the exchange energies that provides an
extra repulsive contribution to the binding energy. Pionic effects are not 
small. Implications of such effects on observables are discussed.
\end{abstract}
\vspace{0.3cm}
\noindent{PACS NUMBERS: 21.65.+f, 24.85.+p, 24.10.Jv, 12.39.-x}


\vspace{1.5cm}
\hspace*{-\parindent}{\bf 1. Introduction}.\hspace*{\parindent}
In a recent series of papers Saito and Thomas (ST)~\cite{ST}\cite{ST-list} 
have developed a model for nuclear matter and finite nuclei in which the quark 
structure of the nucleons is explicitly considered. The model is based 
on an original proposal from Guichon~\cite{Guichon}, and is known as the 
quark-meson coupling model (QMC). The model shares important features with 
the Walecka model (QHD)~\cite{WAL}, such as a relativistic saturation 
mechanism, but the incorporation of explicit quark degrees of freedom has 
nontrivial consequences (for a list of references see Ref.~\cite{ST-list}). 
Recent developments of the model have been made by Jin and Jennings and 
Blunden and Miller~\cite{recent}. For earlier studies within the QMC model see 
Ref.~\cite{earlier}.

Exchange and pionic effects have not been investigated so far in the the QMC 
model. Inclusion of exchange effects (Fock terms) is crucial for assessing 
the effects of pionic degrees of freedom, which are important in view of the 
connection of the pion to the dynamical chiral symmetry breaking of QCD. 
Since the earlier days of QHD, it has been shown~\cite{Chin} that the 
inclusion of Fock terms requires a sizable renormalization of the parameters 
of the model, as compared to the values fixed at the mean field level.
Different parameters of scalar and vector mesons imply for instance in a 
different response of matter to external hadrons immersed in medium. Also, 
the consideration of exchange effects is an initial step towards the 
consideration of short-range correlations (SRC), which have been shown within 
the traditional nuclear many-body theory to play an important role on the 
saturation properties of nuclear matter.

In this paper exchange and pionic effects are studied in the context of a
quark-meson coupling model with constituent quarks. The main motivation for 
using a constituent quark model (CQM) is its simplicity. The CQM has provided 
a good deal of insight into the study of a variety of hadronic phenomena at 
the nuclear scale~\cite{CQM}, and it is natural to expect that it should be 
equally useful for addressing the role of quarks on low-energy properties of 
nuclear matter. In the next section we present the model and discuss 
approximations. In Section~3 we discuss the Hartree approximation and in 
Section~4 we consider the Fock terms and study the effects of the pion. 
Conclusions and future perspectives are presented in Section~5. 

\vspace{0.5cm}
\hspace*{-\parindent}{\bf 2. The model}.\hspace*{\parindent} The mesons are 
described by the Lagrangian density of QHD, including the $\sigma$, $\omega$, 
$\pi$, and $\rho$ mesons. We separate from the $\sigma$, $\omega$, $\rho$ 
field operators their mean-field, classical values $\phi_0$, $\omega_0$ and 
$\vec\rho_0$, and obtain a meson quark-quark interaction from the fluctuating 
part of the fields. The exchange of mesons between quarks is treated in a 
similar fashion to the traditional one-gluon interaction~\cite{deRGG}, with 
the difference that we do not make a nonrelativistic reduction neither of 
the kinetic energy nor of the meson-quark interaction. The exchange energy is the expectation value of the meson 
quark-quark interaction in the state $|\Psi\ra$ of non-overlapping nucleons,
which we write in the form of a  ``Fermi gas of composite nucleons",
\beq
|\Psi\ra = \lim_{N\rightarrow \infty}\,B^{\dag}_{\alpha_1}(\bp_1)\cdots 
B^{\dag}_{\alpha_N}(\bp_N)|0\ra,
\label{Psi}
\eeq 
where the nucleon creation operator is given by 
\beq
B^{\dag}_{\alpha}(\bp) = \frac{1}{\sqrt{3!}}\frac{\epsilon_{ijk}}{\sqrt{3!}}
\frac{T^{abc}_\alpha}{\sqrt{18}}
\int d\bk_1d\bk_2\bk_3\,\Phi_{\bp}(\bk_1,\bk_2,\bk_3)\,
q^{i\dag}_{a}(\bk_1)\,q^{j\dag}_{b}(\bk_2)\,
q^{k\dag}_{c}(\bk_3).
\eeq 
$\Phi_{\bp}$ is the three-quark Fock-space amplitude, where $\bp$ is the 
nucleon center of mass (c.m.) momentum, and $\alpha$ the nucleon 
spin-isospin third components. $T^{abc}_\alpha$ is a spin-isospin 
Clebsch-Gordon coefficient, $a,b,c=(sf)$ are the quark spin-flavor quantum 
numbers and $\epsilon_{ijk}$ is the totally antisymmetric tensor in the color 
indices. The quark creation and annihilation operators satisfy canonical 
anticommutation relations. The largest value of $|\bp|$ in Eq~(\ref{Psi}) is 
the nucleon Fermi momentum~$k_F$, which is related to the nuclear matter 
density as $\rho_N = \gamma k^3_F/6\pi^2$, with $\gamma = 4 (2)$ for nuclear 
(neutron) matter. 

Let us briefly consider the nucleon in free space, and collect some useful
material for later discussions. For simplicity, we neglect gluons and use for 
the confining potential an harmonic oscillator~\cite{CQM}, 
$V(r) =-\lambda^a\lambda^a/4\, C r^2/2$. With a nonrelativistic kinetic 
energy, the problem is exactly soluble. For the ground state,
\beq
\Phi_{\displaystyle{\bp}}(\bk_1,\bk_2,\bk_3) = 
\delta(\bp-\bk_1-\bk_2-\bk_3)\,\left(\frac{3b^4}{\pi^2}\right)^{3/4}\,
e^{-b^2/6\,\sum_{i<j}^3 \displaystyle{(\bk_i-\bk_j)^2}},
\label{Phi}
\eeq
where $b$ is the r.m.s. radius of the nucleon. In case of a relativistic 
kinetic energy, the problem is not exactly soluble. We then use $\Phi$ 
given as above, and treat $b$ as a variational parameter. In this case, the 
ground state mass of the nucleon is given by
\beq
M_N = \left(\frac{3b^2}{2\pi}\right)^{3/2}\int d\bk\,
\sqrt{\bk^2+m^2_q}\,\left[3+\frac{9}{2}\left(b^2\bk^2-1\right)\right]\,
e^{\displaystyle{-3b^2\,\bk^2/2}},
\eeq
where the second term in the square bracket is the energy from the confining 
potential, $3Cb^2$. The constant $C$ was eliminated making use of the 
variational condition on $M_N$. 

In order to calculate the energy density of nuclear matter ${\cal E}$ we need 
the interaction of the quarks with the unconfined, mean meson
fields. We obtain the meson quark-quark interaction in a similar fashion to 
the derivation of the one-gluon interaction in the CQM~\cite{deRGG},
but with the difference that we use Dirac field operators expanded in a 
plane-wave basis, $\psi^i_f(x)= \int d\bk/(2\pi)^{3/2}\,\sum_s u_s(\bk) 
e^{ik{\cdot}x}\,q^{i}_{sf}(\bk)$, with 
\beq
u_s(\bk) = \sqrt{\frac{E^*(\bk)+m^*_q}{2E^*(\bk)}} 
\left(\begin{array}{c}
1 \\
\displaystyle{
\frac{\bsigma\bcdot\bk}{E^*(\bk)+m^*_q} 
}
\end{array}  \right)\chi_s,
\label{spinor}
\eeq
where $E^*(\bk)=\sqrt{\bk^2+m^{*2}_q}$, and 
$m^*_q~\equiv~m_q~-~g_{q\sigma}~\phi_0$ is determined requiring 
stability of the energy density with respect to variations in $\phi_0$. Note
that since antiquarks are neglected, the three-quark nucleon wavefunction,
$<\!\psi(x_1)\psi(x_2)\psi(x_3)|B^{\dag}_{\alpha}(\bp)\!>$, contains only 
positive-energy components. This ensures that the Dirac Hamiltonian is 
bounded from below.  We note however that neglecting antiquarks is not 
entirely consistent since it does not lead to a well defined Hilbert space, 
and should be considered as part of the definition of the model. The use of a 
wave function with a complete Dirac structure requires the use of a different
variational principle to avoid the continuum dissolution problem~\cite{cd}, 
such as the minimax variational principles used in atomic physics for
relativistic many-electron systems~\cite{spv}. In this case, the energy 
functional in the space of variational parameters is a minimum with respect 
to the parameters related to the upper component of the Dirac spinor, and a 
maximum with respect to the ones related to the lower component. Such 
variational calculations have been recently applied to quarks 
models~\cite{fra}, and appear to be particularly suitable for nuclear matter 
calculations like the present one, where a variational calculation seems to 
be imperative. We do not pursue this here, rather reserve it for future 
elaboration, along with the consideration of a proper relativistic structure 
for the confining potential. This last point is still a matter actively 
studied~\cite{szcz-sw}; it is relevant in connection with spin- and 
momentum-dependent effects of the confining interaction, which are not 
considered in the present paper. In the approximation we are using, there is 
no difference between a potential that is the time component of a four-vector 
and a scalar potential in which all spin and momentum-dependent terms are 
neglected in a nonrelativistic expansion. For a wave function with a complete 
Dirac structure however, a confining vector interaction would lead to problems
with the Klein paradox, as is well known~\cite{kp}. One should also note that 
when Fock terms are considered, Eq.~(\ref{spinor}) is not the most general 
spinor basis, since exchange  effects induce a momentum 
dependence for the quark self-energies. In this sense, our approach shares 
similarities with the one of Chin~\cite{Chin} in QHD, where exchange energies 
are evaluated with the mean field self-energies. The same approximation is 
employed in the Dirac-Brueckner approach~\cite{BroMach}. 

For nucleons in medium, we use a gaussian form for $\Phi_{\bp}$ as in 
Eq.~(\ref{Phi}), with $b$ replaced by a $b^*$, and determine $b^*$ 
variationally. Neglecting effects due to the superposition of the quark 
clusters in medium, such as quark exchanges between different nucleons 
(these are discussed in Section~4) and effects thereof~\cite{km}, the energy 
density of nuclear matter is given by
\bea
{\cal E} &=& \gamma \left(\frac{3b^{*2}}{2\pi}\right)^{3/2}
\int_0^{k_F} \frac{d\bp}{(2\pi)^3} \, \int d\bk \left[3\sqrt{\bk^2+m^{*2}_q} 
+ 3\,C\,b^{*2}\right]\, e^{\displaystyle{-3\,b^{*2}\,(\bk-\bp/3)^2/2}}\nn\\
&+& \frac{1}{2}m^2_\sigma \, \phi^2_0 + \frac{1}{2}m^2_\omega \, \omega^2_0
+ \frac{1}{2}m^2_\rho \, \vec{\rho}^2_0 
+ \sum_{p=\sigma\omega\pi\rho} {\cal V}^p_F ,
\label{Eparts}
\eea
where ${\cal V}^p_F$ is the mean field exchange energy 
\beq
{\cal V}^p_F = + \frac{9}{2} \int_0^{k_F}\frac{d\bp}{(2\pi)^3}\,
\int_0^{k_F}\frac{d\bp'}{(2\pi)^3}\,\omega_p(\bp,\bp')\,
e^{\displaystyle{-b^{*2}(\bp'-\bp)^2/3}} ,
\label{VFock}
\eeq
where 
\bea
&&w_p(\bp,\bp') = \left(\frac{3b^{*2}}{2\pi}\right)^{3}
\int d\bk\,d\bq \, e^{\displaystyle{-3b^{*2}/2 [(\bk-\bar{\bk})^2+
(\bq-\bar{\bq})}^2]}\,
\Delta_p(\bt; \delta E^*(\bk,\bt))\,
\frac{T^{(s_4f_4)b_2c_2}_{\alpha}}{\sqrt{18}}\,\nn\\
&&\times\,
[\bar u_{s_4}(\bk)(\Gamma^p)^{f_4f_2} u_{s_2}(\bk+\bt)]\,
\frac{T^{(s_2f_2)b_2c_2}_{\beta}}{\sqrt{18}}\,
\frac{T^{(s_3f_3)b_1c_1}_{\beta}}{\sqrt{18}}\,
[\bar u_{s_3}(\bq)(\Gamma_p)^{f_3f_1} u_{s_1}(\bq-\bt)]\,
\frac{T^{(s_1f_1)b_1c_1}_{\alpha}}{\sqrt{18}},
\label{wpp'}
\eea
where $\bt=\bp'-\bp$, $\bar{\bk}=(\bp-\bt)/3$, $\bar{\bq}=(\bp'+\bt)/3$,
$\delta E^*(\bk,\bt) = E^*(\bk)-E^*(\bk+\bt)$, and the vertices and 
propagators are 
\bea
&&\Gamma^\sigma = i g_{q\sigma},\hspace{0.4cm}
\Gamma^\omega = - i g_{q\omega} \gamma^{\mu},\hspace{0.4cm}
\Gamma^{\pi}_{pv} = \frac{f_{q\pi}}{m_{\pi}} \,{\not\!{q}}\,
\gamma^5 \tau^a ,\hspace{0.4cm}
\Gamma^\rho = \left(- i g_{q\rho} \gamma^{\mu} -
\frac{f_{q\rho}}{2m_q}\sigma^{\mu\nu}q_\nu\right)\tau^a ,
\label{Gammas}\\
&&\Delta_{\sigma,\pi}(\bk; E(\bk)) = 
\frac{1}{E^2(\bk)-\bk^2 - m^2_{\sigma,\pi}} ,
\hspace{0.5cm}
\Delta_{\omega,\rho}(\bk,E(\bk)) = \frac{-1}{E^2(\bk)-\bk^2 - 
m^2_{\omega,\rho}} ,
\eea
where $q=k_3-k_1=k_2-k_4$. We have dropped the term proportional to 
$k^{\mu}k^{\nu}/m^2_{\omega,\rho}$ in the vector potential because of current 
conservation.

As the final step, $m^*_q$ and $b^*$ are determined through the variational 
equations $\partial {\cal E}/{\partial \phi_0} = 0$ and 
$\partial {\cal E}/{\partial b^*}=0$ . Note that $\phi_0$ is a dynamical
variable, and as such can always be determined from the thermodynamic argument
that an isolated system at fixed baryon number and volume (and zero 
temperature) will minimize its energy.

\vspace{0.3cm}
\hspace*{-\parindent}{\bf 3. Hartree Approximation}.\hspace*{\parindent}
The Hartree approximation consists in taking ${\cal V}_F=0$ in 
Eq.~(\ref{Eparts}). The minimization of the energy density with respect to 
$\phi_0$ leads to
\beq
3\,m^*_q = 3\,m_q - 2\, \frac{9g^2_{q\sigma}}{2m^2_\sigma}\,\gamma 
\left(\frac{3b^{*2}}{2\pi}\right)^{3/2}\int_0^{k_F}
\frac{d\bp}{(2\pi)^3} \,\int d\bk \, 
\frac{m^*_q}{\sqrt{\bk^2+m^{*2}_q}}\,e^{\displaystyle{-3\,b^{*2}\,
(\bk-\bp/3)^2/2}}.
\label{gapR}
\eeq
Before discussing the numerical results, we start performing a 
nonrelativistic approximation, with the only purpose of getting insight into 
the problem. All integrals can be done analytically.  Eq.~(\ref{gapR}) becomes
\beq
3\,m^*_q = 3\,m_q - 2\, \frac{9g^2_{q\sigma}}{2m^2_\sigma}\,\left\{ 
\left[ 1 - \frac{3}{5} \, \frac{k_F^2}{2(3m^*_q)^2}\right]
-\frac{1}{2m^{*2}_q\,b^{*2}} \right\}\,\rho_N ,
\label{m*NR}
\eeq
and the energy per nucleon (for symmetrical nuclear matter 
$\vec\rho_0=0$)
\bea
\frac{E}{A}-M_N &=& 3\,\left(m_q^*-m_q\right) 
+ \frac{3}{2m_qb^2}\left( \frac{b^{*2}}{b^2} + 
\frac{m_q}{m^*_q}\,\frac{b^2}{b^{*2}} - 2 \right)
+ \frac{3}{5}\,\frac{k_F^2}{2(3m^*_q)}\nn\\
&+& \frac{1}{4}\,\frac{2m^2_\sigma}{9g^2_{q\sigma}}\,
\left(3m_q-3m^*_q\right)^2\,\frac{1}{\rho_N}
+ \, \frac{9g_{q\omega}^2}{2m^2_{q\omega}}\,\rho_N .
\label{EHNR}
\eea
The variational condition for $b^*$ leads to $b^*/b=(m_q/m^*_q)^{1/4} > 1$, 
since $m^*_q/m_q < 1$. An interesting comparison can be made with QHD,
when the Dirac spinors for the nucleons are expanded to 
${\cal O}(\bp^2/M^2_N)$~\cite{Amore}. The effective mass of the nucleon, 
Eq.~(11) of Ref.~\cite{Amore}, becomes
$M^*_N = M_N -{g^2_\sigma}/{m^2_\sigma}\,(1- {3k^2_F}/{10M^{*2}_N} \,)\rho_N.$
Comparing this with Eq.~(\ref{m*NR}) above, it becomes evident that the effect
of internal structure of the nucleon is the term $1/2m^{*2}_q\,b^{*2}$ in 
Eq.~(\ref{m*NR}). This term comes from the kinetic energy of the quarks in the
nucleon, and is of the opposite sign to the QHD term. The internal motion of 
the quarks provides an extra repulsion and acts as a repulsive field in 
medium. 

We next consider the numerical solutions of the relativistic equations. 
In this case the integrals in Eqs.~(\ref{Eparts})(\ref{VFock})(\ref{wpp'}) 
and (\ref{gapR}) must be performed numerically. The determination of $b^*$ 
must also be done numerically. We fix the coupling constants such as to 
obtain a stable minimum of $E/N-M_N \simeq -15.75$~MeV at 
$k_F \simeq 1.36$~fm$^{-1}$. We use standard 
values for the masses, namely~$m_q=350$~MeV,~$m_\sigma=550$~MeV, and 
$m_\omega=783$~MeV. The parameter set is shown in Table~I, where we show 
results for $b=0.6$~fm and $b=0.7$~fm. 

In Fig.~1 we plot $E/N-M_N$ as a function of $k_F$ for different model
parameters of Table~I. In Fig.~1(a), the QHD mean-field result with parameter 
set~(1) in Table~I, is the long-dashed line. Our Hartree result for 
$b=0.6$~fm, parameter set~(2), is the long-short--dashed line. Inspection of 
Fig.~1(a) reveals that our Hartree result provides a nuclear compressibility, 
$K$, that is smaller than the one obtained in QHD. As seen in Table~I, the 
values for $K$ for $b=0.6$~fm and $b=0.7$~fm are considerable smaller than 
the value in QHD (MFT), $K=$~545~MeV. The qualitative feature of the results 
is that the effective repulsion provided by the internal degrees of freedom 
of the nucleons increases as $b$ decreases. 

The original QMC model of ST predicted a much smaller mean $\omega$ field as 
predicted in finite-density QCD sum rules and relativistic nuclear 
phenomenology~\cite{vecf}. On the other hand, when the correction due to the 
c.m. in the ST model is taken to be independent of the applied field, the 
discrepancies are significantly reduced. Our value of $g_{\omega}$ is very 
similar to the new value in the ST model~\cite{ST-list}.

\vspace{0.3cm}
\hspace*{-\parindent}{\bf 3. Exchange and pionic contributions}.
\hspace*{\parindent}
The evaluation of the Fock energy involves the summation over the spin-isospin
quantum numbers of the quarks and the evaluation of a seven-dimensional 
numerical integration in Eqs.~(\ref{VFock})(\ref{wpp'}). The spin-isospin sum 
can be easily performed using the ``substitution rules"~\cite{lst}, which 
allow to express quark spin-isospin Pauli matrices in terms of nucleon 
spin-isospin Pauli 
matrices. Regarding the multidimensional integral, although it can be carried 
out using a Monte Carlo integrator, we notice that the integrands of the $\bk$
and $\bq$ integrals in Eq.~(\ref{wpp'}) are concentrated around the values of 
$\bar{\bk}$ and $\bar{\bq}$, respectively . In view of this, we use 
$\bar{\bk}$ for $\bk$ and $\bar{\bq}$ for $\bq$ in the spinors and in
the meson propagators and integrate analytically over $\bq$ and $\bk$.
What remains is a three-dimensional integral, which has to be performed
numerically. The minimization of the energy density with respect to $\phi_0$ 
and $b^*$ is also performed numerically.

The new values of $g_{\sigma}$ and $g_{\omega}$ are shown in Table~I.
The readjustment of $g_{\omega}$ is relatively small, of the order of $-$10\%, 
and of $g_{\sigma}$ is of the order of $+$40\%. As in QHD, the effect of the 
Fock terms is repulsive, as seen in Fig.~1(a), where the curve with 
short-dashes is calculated with the Hartree parameter set~(2), including the 
exchange terms from the $\sigma$ and $\omega$ mesons. However, the internal 
structure of the nucleon has an extra, nontrivial effect, which can be 
qualitatively understood through a nonrelativistic reduction of the Fock 
energy from the $\sigma$ meson
\bea
{\cal V}^s_F &=& + \frac{1}{2}\,9g^2_{q\sigma}\,\gamma
\int_0^{k_F}\frac{d\bp}{(2\pi)^3}\int_0^{k_F}\frac{d\bp'}{(2\pi)^3}\,
\frac{e^{\displaystyle{-b^{*2}(\bp-\bp')/3}}}{(\bp-\bp')^2+m^2_s}\,
\Biggl\{\left[1-\frac{(\bp+\bp')^2}{4(3m^*_q)^2}\right]-\frac{1}{m^{*2}_q b^2}
\Biggr\}.
\eea
The first term in square brackets is the nonrelativistic reduction of the 
exchange energy in QHD~\cite{Chin}. The term $1/m^{*2}_q b^2$ represents the 
nontrivial effect of the internal structure of the nucleon, which is of the 
opposite sign to the QHD term, and is responsible for the enhancement of 
$g_{\sigma}$. This term is equivalent to the $\sigma$ dependence of the 
$g_{\sigma N}$ coupling constant in the original QMC work~\cite{ST}. A 
similar, but smaller effect is seen in the contribution from the $\omega$.
The enhancement of $g_{\sigma}$ is of interest for a good description of the 
spin-orbit splittings in finite nuclei~\cite{ST-list}.

The $\pi$-quark coupling constant is determined~\cite{lst} by comparing the 
asymptotic behavior of the nucleon-nucleon one-pion-exchange potentials at 
the quark level and for point nucleons:
$f^2_{q\pi}=(9/25)\,f^2_{N\pi}\,\exp(-m^2_\pi b^2/2)$, with $f^2_{N\pi}=
g^2_{N\pi}(m^2_{\pi}/4M^2_N)=0.98$, $m_\pi~=~138$~MeV and 
$g^2_{N\pi}/4\pi=14.4$. The effect of the $\pi$ is not small, although somewhat
smaller than the exchange effects from the $\sigma$ and $\omega$ mesons. The 
dash-dotted curve is calculated with the H parameter set~(2), and includes
the exchange effects of the $\sigma$ and $\omega$. The new readjustment of 
$g_{\sigma}$ and $g_{\omega}$ to saturate nuclear matter at the right place
is relatively small, as seen in Table~I, entry (5). 

Because of the large tensor coupling of the $\rho$ meson, which 
is of the comparable magnitude to the $\pi$ coupling, we also include 
its effects in the present calculation. The $\rho$-quark couplings are fixed 
as in the case of the $\pi$. For comparison with the results of 
Ref.~\cite{franc}, we use $g^2_{N\rho}/4\pi=0.55$, 
$f_{N\rho}/g_{N\rho}=3.7$, and $m_\rho=770$~MeV. The effect of the $\rho$ is 
also repulsive. The new values for $g_{\sigma}$ and $g_{\omega}$ are 
given by the parameter set~(6) in Table~I. The solid curve in Fig.~1(a) is the
result of including all mesons.

As in QHD~\cite{WAL}, the effect of the $\pi$ and the tensor coupling
of the $\rho$ are repulsive. This is due to the zero-range (ZR) components 
in the interaction, which would not contribute if short-range correlations 
are taken into account. 
In order to obtain a more reasonable estimate of the 
effects of $\pi$ and $\rho$, we follow Ref.~\cite{franc} and remove this 
component from the interaction. The effect provided by the attractive $\pi$
is of the order of $10$~MeV, similar do the effect in the QHD calculation of
Ref.~\cite{franc}. The long-dashed curve in Fig.~1(b) is calculated with the 
H parameter set~(2) and the ZR component of the $\pi$ removed, and includes 
the exchange terms of $\sigma$ and $\omega$ mesons. The long-short--dashed 
curve in Fig.~1(b) is calculated with the H parameter set~(2) and the ZR
components from both the $\pi$ and the $\rho$ interactions removed. The net 
attraction provided by the $\pi$ and $\rho$ is of the order of $24$~MeV per 
nucleon.In the calculation of 
Ref.~\cite{franc}, for the same value of $f_{N\rho}/g_{N\rho}=3.7$, the net 
attraction is of the order of $26$~MeV per nucleon. 

To conclude, we note that the exchange effects considered here
arise from the Pauli principle at the nucleon level, i.e. quark-exchange (QE)
between different nucleons were neglected. QE effects are expected to be small
at low densities, but for for nucleons with $b~=~0.6$~fm in a medium at the 
normal nuclear density, they might have interesting consequences. In 
particular, they might be of interest for replacing $\omega$ exchange as the 
main source of the short-range repulsion, in view of possible 
conceptual difficulties with the coupling of an extended meson to quarks 
inside the nucleons~\cite{ST}~\cite{Guichon}. There is an extensive 
literature~\cite{clust} in this subject and it is out of the scope of the 
present paper to review the subject here. The present model can 
be naturally extended to include QE effects following the 
idea~\cite{lst} that the innermost, short-ranged part of 
the NN force is generated by QE with pion coupling, and the outer 
part is generated by meson coupling to the nucleon. In such a picture, there
is no danger for double counting by taking simultaneously QE and 
$\omega$ exchange, since the $\omega$ coupling to the nucleon is cut-off by 
the nucleon form factor. We use the approach of 
Ref.~\cite{FT} to derive an effective NN interaction from pion-quark exchange.
The resulting NN interaction is non-local and gives scattering phase-shifts 
numerically similar to the ones obtained with the resonating group 
method~\cite{clust}. After a long calculation, with the extensive use of the 
substitution rules of Ref.~\cite{lst}, the contribution of QE to the energy 
density of symmetrical nuclear matter can be written as 
\beq
{\cal V}^{exch}_q = \frac{f_{q\pi}^2}{3m^2_\pi}
\int_0^{k_F}\frac{d\bp}{(2\pi)^3}\,
\int_0^{k_F}\frac{d\bp'}{(2\pi)^3}\,\sum^{5}_{k=1}A_k 
\;e^{\displaystyle{-a_k\,b^{*2}\,(\bp-\bp')^2}} ,
\eeq
with $A_1=8\,(3/4)^{3/2},\;a_1=1/12$, 
$A_2 = 54,\;a_2 = 0$, 
$A_3 = 120\,(12/11)^{3/2},\; a_3 = 2/33$,
$A_4 =-44/3,\; a_4 = 1/3$, 
$A_5 =-272/3\,(12/11)^{3/2},\;a_5 = 8/33$. 
Here, only the contribution of the zero-range part of the 
pion-exchange is shown. The contribution from the long-range part is 
numerically negligible, as also found in Ref.~\cite{lst}, where a 
nonrelativistic form for the pion-exchange was used. The new coupling 
constants (we do not consider $\rho$ coupling) are $g^2_\sigma = 173.45$ 
and $g^2_\omega = 56.2$. The compressibility is $K=375$~MeV and 
$b^*/b=1.07$. Note that this value of $g^2_\omega$ is very close to the 
quark-model SU(6) symmetry prediction ${g^2_\omega}/4\pi~\approx~9 
{g^2_\rho}/4\pi~=~9\,\times 0.55$, which is much smaller than the values used 
in QHD and in one-boson-exchange models~\cite{BroMach}. We have also checked 
the effect of quark exchange with $\sigma$ coupling; the effect amounts to a 
new readjustment of~$g_\sigma$, with no important consequences for the values 
of $K$ and $b^*/b$.

\vspace{0.3cm}
\hspace*{-\parindent}{\bf 4. Conclusions and Future Perspectives}.
\hspace*{\parindent}
The consideration of exchange effects in QMC-type of models is the 
initial step towards the implementation of chiral symmetry and short-range
correlations. Important consequences are expected when 
both elements are put together, as recently shown by Banerjee and 
Tjon~\cite{BT} in a study using density-dependent meson masses and couplings. 
We have shown here that the coupling of 
mesons directly to the quarks has important consequences on the exchange 
energy, implying in a sizable readjustment of the meson-quark coupling 
constants. A readjustment of coupling constants is not meaningless, since 
different coupling constants imply in different in-medium spin-spin, 
spin-orbit and tensor forces and therefore different responses will be 
experienced by external hadrons immersed in medium.

An interesting new direction is the replacement of $\omega$-coupling by 
nucleon overlap at short distances. The idea of describing the very 
short-range part of the NN interaction by quark-exchange has long been 
discussed and we have shown that an interesting consequence is a large 
reduction of the value of $g_\omega$. Another interesting aspect that is 
expected to influence the saturation mechanism is the inclusion of the 
low-lying nucleon resonances. The present model has the potentiality to 
incorporate such effects, since with the values 
$\hbar\omega = 1/m_q^2b^2 = 174$~MeV and $g_{q\pi}$ used here it is possible 
to obtain a reasonably good description of the low-mass spectrum of the 
nucleon, including the Roper resonance~\cite{GR}.

An important and necessary next step is to improve on the Lorentz 
transformation properties of QMC-type of models, based on bag or 
semi-relativistic constituent quark models, a problem that is intimately 
related to the projection of the c.m. motion. This is
because low energy theorems~\cite{LET} based on Lorentz invariance impose 
constraints on the single-nucleon energy in matter which have a 
direct impact on the saturation properties of nuclear matter. 

\vspace{0.3cm}
\hspace*{-\parindent}{\bf Acknowledgments}.\hspace*{\parindent}
GK thanks R.~Brockmann for discussions and suggestions. GK~also
thanks G.A.~Miller and A.W.~Thomas for discussions on the QMC model during the
program ``Quark and gluon structure of nucleons and nuclei" of the National 
Institute for Nuclear Theory in Seattle. This work 
was supported in part by the Alexander von Humboldt Foundation (Germany) and 
FAPESP and CNPq (Brazil).

\newpage

\vspace{0.5cm}
\noindent
TABLE I: Model parameters and results for the compressibility $K$ and 
$b^*/b$. The parameters $g_{\omega}=3g_{q\sigma}$ and 
$g_{\sigma}=3g_{q\sigma}$ are adjusted to $E/A-M_N \simeq -15.75$~MeV
at $k_F \simeq $ 1.36 fm$^{-1}$. 

\begin{center}
\begin{tabular}{lddd|dd}
\tableline\tableline
Model&$\;\;b$ (fm)$\;\;$&$\;\;\;g_\sigma^2\;\;\;$ &$\;\;\;g_\omega^2\;\;\;$
&$\;\;K$ (MeV)$\;\;$&$\;\;b^*/b\;\;$ \\
\tableline
(1) QHD (MFT)                        &  -  & 105.96  & 161.10    & 545  &    
 -     \\
(2) H                                & 0.6 & 117.00  &  90.27    & 330 & 
1.045  \\
(3) H                                & 0.7 & 112.10  &  98.80    & 340 & 
1.05   \\
(4) H+EXCH ($\sigma+\omega$)         & 0.6 & 152.89  &  88.43    & 370 & 
1.06   \\
(5) H+EXCH ($\sigma+\omega+\pi)$     & 0.6 & 165.60  &  90.90    & 362 & 
1.07   \\
(6) H+EXCH ($\sigma+\omega+\pi+\rho)$    
                                     & 0.6 & 165.66  &  89.80    & 351 & 
1.07   \\
(7) H+EXCH ($\sigma+\omega+\pi_{\text{att}})$     
                                     & 0.6 & 147.44  &  98.75    & 353 &
1.06 \\
(8) H+EXCH ($\sigma+\omega+\pi_{\text{att}}+\rho_{\text{att}})$    
                                     & 0.6 & 137.71  &  101.50   & 340 & 
1.05   \\
\tableline\tableline
\end{tabular}
\end{center}
\vspace{0.5cm}

\vspace{1.0cm}
\epsfxsize=14.0cm
\centerline{\epsfbox{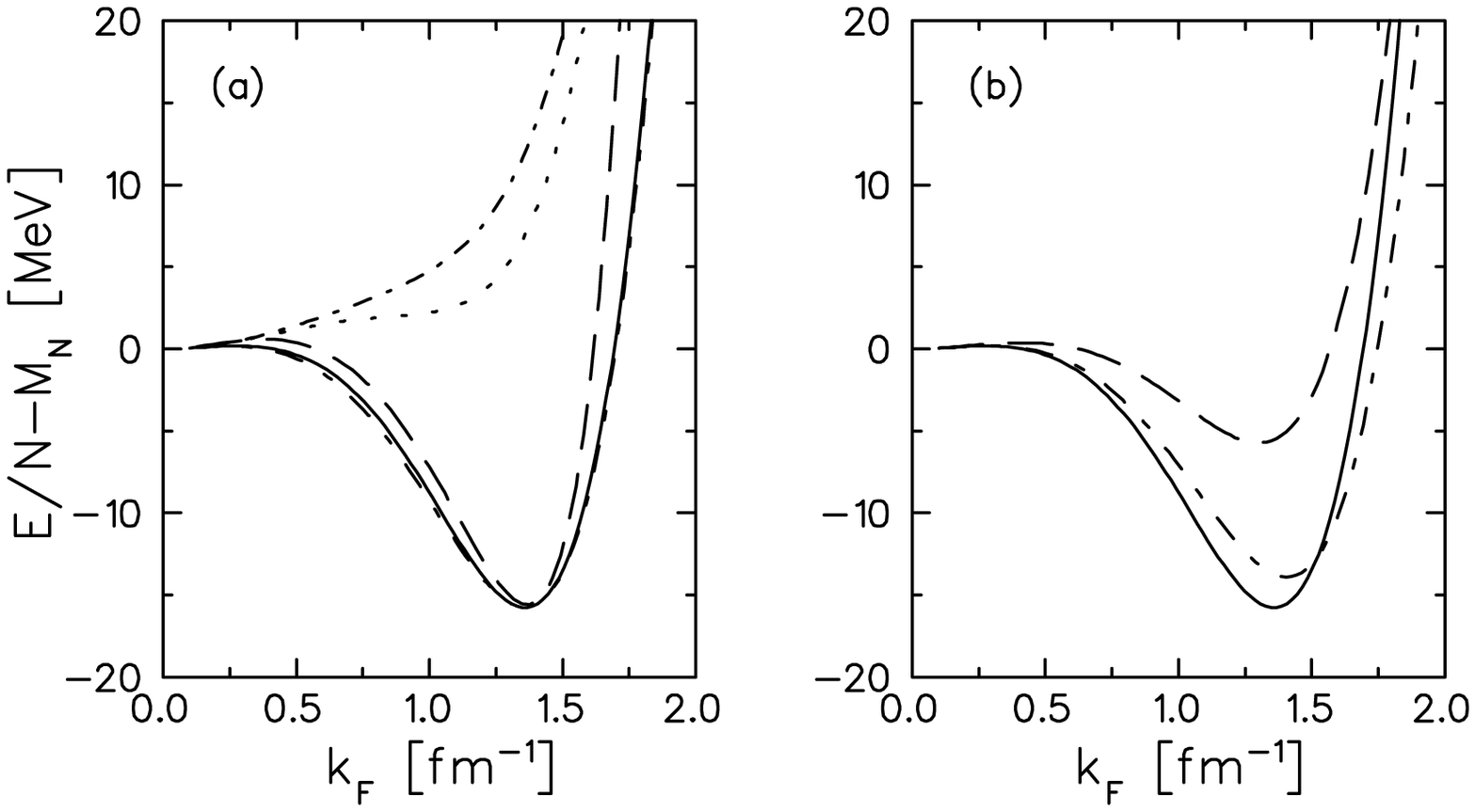}}

\noindent
FIGURE 1. The total energy per particle as a function of the Fermi momentum.
(a) The long-dashed curve corresponds to QHD (MFT). The long-short--dashed 
curve is our Hartree result, parameter set (2) in Table~I, and the solid curve
is for parameter set (6). The short-dashed (dot-dashed) curve is calculated 
with the H couplings (2) including exchange terms of $\sigma+\omega$ 
($\sigma+\omega+\pi$). (b) The solid curve is for parameter set (8), with the
zero-range parts of $\pi$ and $\rho$ removed. The long-dashed 
(dash-dotted) is calculated with the H couplings~(2) and exchange
terms of $\sigma+\omega+\pi_{\text{att}}$ 
($\sigma+\omega+\pi_{\text{att}}+\rho_{\text{att}}$). All curves are for 
$b=0.6$~fm.
\vspace{0.5cm}

\end{document}